\documentclass[sigconf]{acmart}

\AtBeginDocument{%
  \providecommand\BibTeX{{%
    \normalfont B\kern-0.5em{\scshape i\kern-0.25em b}\kern-0.8em\TeX}}}

\usepackage{multirow} 
\usepackage{makecell} 
\usepackage{todonotes}
\usepackage{appendix}
\usepackage{stfloats}

\copyrightyear{2021}
\acmYear{2021}
\setcopyright{acmlicensed}\acmConference[CHI '21]{CHI Conference on Human Factors in Computing Systems}{May 8--13, 2021}{Yokohama, Japan}
\acmBooktitle{CHI Conference on Human Factors in Computing Systems (CHI '21), May 8--13, 2021, Yokohama, Japan}
\acmPrice{15.00}
\acmDOI{10.1145/3411764.3445432}
\acmISBN{978-1-4503-8096-6/21/05}

\newcommand\revision[1]{{\color{black} #1}}

\begin{document}

\title{``Brilliant AI Doctor'' in Rural China: Tensions and Challenges in AI-Powered CDSS Deployment}

\author{Dakuo Wang}
\authornote{Both authors contributed equally to this research.}
\email{dakuo.wang@ibm.com}
\affiliation{
    \institution{IBM Research}
    \country{USA}
}

\author{Liuping Wang}
\authornotemark[1]
\affiliation{
\institution{Institute of Software, Chinese Academy of Sciences}
}
\affiliation{
    \institution{University of Chinese Academy of Sciences}
    \country{China}
}

\author{Zhan Zhang}
\affiliation{
    \institution{Pace University}
    \country{USA}
}
  
\author{Ding Wang}
\affiliation{
    \institution{Microsoft Research India}
    \country{India}
}
  
\author{Haiyi Zhu}
\affiliation{
    \institution{Carnegie Mellon University}
    \country{USA}
}

\author{Yvonne Gao}
\affiliation{
    \institution{The State University of New York at Stony Brook}
    \country{USA}
}

\author{Xiangmin Fan}
\affiliation{
    \institution{Institute of Software, Chinese Academy of Sciences}
    \country{China}
}
  
\author{Feng Tian}
\affiliation{%
    \institution{Institute of Software, Chinese Academy of Sciences}
    \country{China}
}

\renewcommand{\shortauthors}{Wang and Wang, et al.}
\renewcommand{\shorttitle}{``Brilliant AI Doctor'' in Rural China: Tensions and Challenges in AI-Powered CDSS Deployment}

\begin{abstract}
Artificial intelligence (AI) technology has been increasingly used in the implementation of advanced Clinical Decision Support Systems (CDSS). Research demonstrated the potential usefulness of AI-powered CDSS (AI-CDSS) in clinical decision making scenarios. However, post-adoption user perception and experience remain understudied, especially in developing countries. Through observations and interviews with 22 clinicians from 6 rural clinics in China, this paper reports the various tensions between the design of an AI-CDSS system (``Brilliant Doctor'') and the rural clinical context, such as the misalignment with local context and workflow, the technical limitations and usability barriers, as well as issues related to transparency and trustworthiness of AI-CDSS. Despite these tensions, all participants expressed positive attitudes toward the future of AI-CDSS, especially acting as ``a doctor's AI assistant'' to realize a Human-AI Collaboration future in clinical settings. Finally we draw on our findings to discuss implications for designing AI-CDSS interventions for rural clinical contexts in developing countries.
\end{abstract}



\maketitle

\section{Introduction}
Artificial intelligence (AI) technologies are increasingly being developed for a range of clinical circumstances ~\cite{doctorbot,esteva2017dermatologist,cass,erickson2017machine}. In particular, using AI techniques to help clinicians make fast and accurate diagnostic decisions is gaining momentum~\cite{cai2019human}. This application of AI falls under the category of a clinical decision support system (CDSS), which takes patient data as input and generates patient-specific advice ~\cite{moja2014effectiveness}. Unlike `traditional' CDSS that relies on expert-defined heuristics, modern AI-driven CDSS interventions utilize state-of-the-art AI algorithms (e.g., deep neural networks and knowledge graph) \revision{to train tremendously large datasets retrieved from electronic health record (EHR) systems~\cite{middleton2016clinical}.} This new generation of CDSS can replicate the natural flow of clinical reasoning at the point-of-care so that patient-specific recommendations are generated to augment clinician decision-making ~\cite{moja2010navigators}. In this paper, we use \textbf{AI-CDSS} to collectively represent the group of CDSS systems that are powered by advanced AI algorithms~\cite{stone2016one}. 

Despite the seminal studies on the technical aspects of applying AI to enhance the performance and effectiveness of CDSS (e.g., ~\cite{cai2019human,chen2013my,Zhang2013,yang2019unremarkable,singh2017hci,coyle2009clinical,kulp2018design,sultanum2018more,cai2019human,zhang2018medical}), very little is published on the usage of AI-CDSS in practice ~\cite{brufau2019lesson,jiang2017artificial}. This research gap is critical because researchers need an in-depth understanding of how to appropriately implement AI-CDSS to increase user acceptance~\cite{doctorbot} and system uptake ~\cite{shortliffe2018clinical}. Furthermore, the unique characteristics of AI may hinder AI-CDSS adoption and thus, demanding further studies ~\cite{brufau2019lesson}. For example, clinicians may resist accepting AI-CDSS because of their perceptions that AI will replace them from their jobs ~\cite{gong2019influence}. Another major barrier is the interpretability and transparency of AI, that is, how AI-CDSS works to generate the recommendation inside the algorithm ``black box''~\cite{weidele2020autoaiviz} remains opaque to clinicians ~\cite{shortliffe2018clinical}. Given these barriers, there is a general consensus that social-technical context must be considered when developing and implementing AI-CDSS~\cite{strickland2019ibm, peiris2011new}, and evaluation of these systems in real-world clinical context is essential for identifying barriers that inhibit successful implementation ~\cite{magrabi2019artificial, moja2014barriers}.

In addition to the lack of understanding of AI-CDSS usage and intergration in practice, the second prominent research gap is that the majority of research around AI-CDSS is situated in the context of developed countries, where the \revision{implementation of CDSS has been far earlier} than the developing countries ~\cite{sambasivan2012intention}. Despite some developing countries have increased their investment in health information technologies (HITs) ~\cite{li2019efficiency}, these countries often face much greater challenges compared to the developed countries in implementing AI-CDSS interventions ~\cite{tomasi2004health}. For example, CDSS normally requires a robust EHR system for gathering patient data to generate accurate diagnostic recommendations, but the implementation of EHR systems is still very limited in many developing countries ~\cite{sambasivan2012intention}. Furthermore, clinics in the rural areas of developing countries usually have even less access to well-integrated HITs. The clinicians in these rural clinics also lack training opportunities \revision{to learn how to use sophisticated HITs}, which further limits the adoption of AI-CDSS systems~\cite{li2019efficiency}. When it comes to the AI era, these social and cultural factors can influence the success and outcome of AI-CDSS introduction in developing countries ~\cite{moja2014barriers}; \revision{therefore, these factors need to be carefully examined to facilitate seamless implementation of AI-CDSS in such settings ~\cite{coiera2015guide}.}

In this work, we join the research effort of enriching empirical understandings of clinicians' perception and usage of AI-CDSS in developing countries. To that end, we conducted a fieldwork in clinics in a rural region of China, where an AI-CDSS system (``Brilliant Doctor'') was just deployed over six months before this study took place~\footnote{We have no affiliation with the company that developed the system nor with the government agency that decided to deploy the system. We read the news~\cite{AIhealthcare}, then we directly reached out to the clinics to interview their usage experience of the system. Shortly after our study was completed, the Brilliant Doctor system was acquired by Baidu Inc. https://01.baidu.com/cdss-basic.html}. As the most populated country in the world, hundreds of millions of Chinese still live in rural areas and they often have low income and limited access to healthcare resources  ~\cite{liu2007rural,yip2008chinese}. We believe the findings of this work can shed lights on how to design and implement AI-CDSS systems for similar rural clinical contexts in other developing countries. 

Through this study, we make the following contributions: 
\begin{itemize}
  \item A field work examining how AI-CDSS is used in rural clinics of developing countries, and how its usage is shaped by social, cultural, and contextual factors;
  \item An empirical understanding of clinicians' perceptions and facing challenges with regard to the use of AI-CDSS in practice;
  \item Design implications and recommendations for increasing the uptake and utilization of AI-CDSS to realize a \revision{Clinician-AI Collaboration paradigm} in rural areas of developing countries.
\end{itemize}

\section{Background and Related Work}
In this section, we will review multiple strands of literature to situate our work, including CDSS interventions, issues in the use of AI-CDSS, and implementations of HITs in developing countries. 

\subsection{Clinical Decision Support Systems}
CDSS interventions have been playing an important role in enhancing medical decision-making ~\cite{osheroff2007roadmap}, such as assisting clinicians in diagnosing patients ~\cite{mosquera2014computer}, suggesting medication and treatment options ~\cite{zamora2013pilot}, and flagging potential adverse drug reactions and allergies ~\cite{moxey2010computerized}. \revision{Over the past three decades}, we have seen a dramatic evolution of CDSS in many ways. Early CDSS systems such as QMR ~\cite{miller1986internist}, DXplain ~\cite{barnett1987dxplain}, and Meditel ~\cite{worley1990meditel} are all rule-based or heuristic-based systems that heavily relied on human experts to manually craft knowledge-bases and heuristics ~\cite{wright2009clinical,berner2009clinical}. While the literature consistently demonstrated the potential of these systems to support evidence-based practice, they have inherent limitations. For example, they generally have limited number of disease states or conditions and suffer from the difficulty of maintaining the rule knowledge base in an up-to-date format ~\cite{middleton2016clinical}. In addition, they lack the appropriate capability to interpret ambiguous patient cases or unseen scenarios ~\cite{sittig2008grand,shiffman2005guideline, kong2008clinical}. These challenges still persist to the present day in many CDSS systems, leading to non-optimal performance and low adoption rate ~\cite{berner1994performance}. 

In recent years, with the advent of artificial intelligence techniques, such as neural network models ~\cite{hansen1990neural} and knowledge graph~\cite{wang2014knowledge}, there has been an explosion in the development of AI technologies to improve CDSS performance and accuracy ~\cite{patel2009coming}. These new AI methods, combined with the availability of large clinical datasets, revolutionize CDSS interventions from rule-based approaches toward a more ``big data'' and numeric-based approach ~\cite{lindsey2018deep, curioni2017new, tian2019can}. It is expected that these new generations of CDSS that utilize AI technologies may soon outpace the use of ``traditional'' methods for clinical decision support rule generation ~\cite{middleton2016clinical}.

However, these AI-driven CDSS systems are still in developmental phase and very few systems have been deployed in clinical settings. Due to the lack of large-scale deployment, it remains unclear \revision{about how such systems are perceived and used} by their intended users (e.g., clinicians), and what barriers and challenges exist in their deployment. Gaining an in-depth, empirical understanding of these aspects will help the designers and developers of AI-CDSS to better understand and address potential issues in fitting AI-CDSS into clinical practice. Our study contributes to bridging this knowledge gap by examining the use of an AI-CDSS system in real world settings.

\subsection{Issues in the Implementation of AI-CDSS}
Despite consistent findings demonstrating the potential of CDSS to support evidence-based practice and improve patient outcomes, the literature has shown that AI-CDSS applications may face difficulties when integrating into the real-world practice ~\cite{ridderikhoff1999afraid}. Many barriers have been identified for the low usage of CDSS interventions, including too frequent or false alarms, poor human interface design, interference with established workflow, time pressures, and inadequate training ~\cite{patterson2005identifying,zheng2005understanding,rousseau2003practice, short2004barriers, saleem2005exploring}. These prior studies provide insights with regard to primary reasons for the failure of CDSS implementation. However, given the increasing complexity of the constantly developing AI technologies, the implementation of AI-CDSS may have unique and unexpected barriers that need careful and thorough investigation, which otherwise could easily leave the new AI-CDSS system unused (just as the old ones), which is one of the biggest issues faced by the developers and designers of HITs~\cite{strickland2019ibm,sambasivan2012intention}. 

One major barrier of AI that is especially relevant to AI-CDSS is the \textit{``black box''} issue ~\cite{brufau2019lesson,weidele2020autoaiviz}. That is, the inner workings of the algorithm that include the rationale for how the predictions are generated and how different clinical features are considered in the model remain opaque to the medical professionals. This challenge impedes users developing appropriate understanding and trust of the output yielded by AI systems ~\cite{wang2019humanai,drozdal2020trust,stone2016one,cai2019human}. One recent research thread, \textit{Interactive Machine Learning}~\cite{fails2003interactive}, is a promising solution to the ``black box'' issue of AI systems. Interactive machine learning often relies on advanced visualization and interaction techniques (e.g.,~\cite{drozdal2020trust,weidele2020autoaiviz,modellineupper}), allowing users to monitor, interpret, and intervene the machine learning process so that they can establish trust toward the algorithm outputs. In a similar vein, a group of Google researchers recently built an experimental visualization system to help physicians understand the AI interpretations of  breast cancer imaging data~\cite{cai2019human}.

Another aspect of AI-CDSS that must be carefully navigated is medical professionals' feeling and perception of AI ~\cite{brufau2019lesson}. It is a general consensus that medical professionals' attitudes play an important role in the adoption of CDSS ~\cite{moja2014barriers}. \revision{But recent studies documented medical professionals' concerns about the possibility of AI replacing their jobs as AI becomes more powerful and takes more initiatives ~\cite{gong2019influence}.} This is related to the ``professional autonomy'' concept---the clinicians need guarantee to freely operate their professional judgement and decision-making in patient care without any interference ~\cite{ten2000re}. Prior work has suggested that the perceived threat of a technology to professional autonomy and the degree to which clinicians consider the use of a particular system decreases their control over decision making are salient factors that could directly affect clinicians' willingness to use that technology ~\cite{walter2008physician,sambasivan2012intention,friedberg2014factors}.

Both the ``black box'' concern and the ``threat to professional autonomy'' concern highlight the emerging need of re-considering the relationship between the agency of human users and the agency of the more advanced, yet less transparent, AI technologies (e.g.,~\cite{park_2019,jung2011portal,wang2019humanai,jacques2019conversational}). In a CHI'19 workshop entitled \textit{``Human-AI Collaboration in Healthcare''}, researchers discussed challenges of this new paradigm, including the human bias and inequity of healthcare resources that are amplified by the AI algorithm~\cite{park_2019}. \revision{This workshop discussion echos the broad Human-AI Collaboration research agenda happening in various application domains: education~\cite{xu161same}, data science~\cite{wang2019humanai,autods}, and workplace~\cite{shamekhi2018face,liao2018all}.} One salient research question arose across all these domains is whether and how we can build cooperative AI systems that human users and AI can work as partners? \revision{Our findings from this paper can also provide timely insights for this critical research question.}


In this work, we join the effort of ongoing work ~\cite{cai2019human,stone2016one, yang2019unremarkable,park_2019,beede2020human} to design effective, interpretable, user-centered, and context-aware AI-CDSS systems. As the beginning step of a larger scale research project, we first conducted a field study to examine the use of a recently implemented AI-CDSS system in practice. More specifically, we aim to provide an empirical understanding of how clinicians are using this system in their day-to-day work practices and what barriers exist in the use of emerging AI-CDSS applications. The findings from our study can shed lights on new research directions for both HCI and AI communities, for example, how to design and develop locally-adapted and clinician-accepted AI-powered health technologies.

\subsection{Healthcare Information Technologies in Developing Countries}
There has been growing investment in the implementation of HITs, including CDSS, in developed countries. For example, in 2009, the Health Information Technology for Economic and Clinical Health act was passed with \$27 billion funding in the United States~\cite{sharma2016impact}. This trend has been driven by the premise that these systems can help improve clinician performance and enhance the quality of healthcare ~\cite{ash2012recommended}. However, such heavy investment and the mere provision of the technology does not guarantee its uptake. A recent study involving outpatient physicians in U.S. shows that CDSS is being used only in 17\% of the patient visits ~\cite{romano2011electronic}. Also, clinicians often reject the recommendations of CDSS, ignoring up to 96\% of its alerts ~\cite{moxey2010computerized}. Even more concerning is that the regional differences in the adoption of HITs have been problematic in U.S. ~\cite{hsiao2012use}, where under-served and rural areas are especially affected by less access to HITs ~\cite{king2013geographic}. 

The uptake of advanced HITs is not great in rural areas of developed countries, but what about the situations in the global south? Developed countries have adopted \revision{HITs and built healthcare system infrastructure far earlier than the developing countries~\cite{sambasivan2012intention};} thus, it is not hard to picture that the developing countries face a much greater challenge in implementing CDSS systems in practice ~\cite{tomasi2004health}. For example, CDSS depends on EHR to gather relevant patient data, but the implementation of EHR in developing countries is limited ~\cite{fraser2005implementing}. Shu et al. ~\cite{shu2014ehr} investigated EHR adoption in 848 tertiary hospitals in China, and found that most hospitals have only adopted a very basic EHR system and thus, the patient records need to be printed out to be shared across hospital departments. To address this issue, many hospitals in developing countries have turned to personal digital communication tools for assistance. For instance, Karusala et al. \cite{karusala2020making} described how WhatsApp is adopted in an Indian hospital to facilitate the organizational communication by nurses while \cite{wang2020please} reported a case of a Chinese clinic using WeChat as its patient-facing communication channel. \revision{Critical medical information is discussed and shared between patients and clinicians using these personal chat tools \cite{cass}. However, the challenge is that these discussions (including shared patient information) were rarely transmitted into the EHR systems for formal documentation, causing loss of patient data. The status quo of the implementation of EHR systems largely contributes to the low usage of CDSS in developing countries.} 

\revision{Other cited barriers of implementing CDSS in developing countries included insufficient computer literacy of clinicians and the significant cost associated with purchasing CDSS applications ~\cite{coiera2015guide}.} However, there has been very limited research focusing on the social, cultural, and contextual factors in this context. \revision{More recently, researchers have begun emphasizing the importance of incorporating local stakeholders' concerns and organizational contexts in developing acceptable and successful AI systems ~\cite{zhu2018value, cheng2019explaining}.}
Therefore, our work contributes to this line of work by \revision{examining the interrelationship between the local practice culture and the use of AI-CDSS in rural areas of China through a social-technical lens.} In particular, we are interested in how clinicians make use of and think about AI-CDSS systems in rural clinics of developing countries.

\section{Method}
In early 2019, an AI-CDSS system,  ``Brilliant Doctor'', was introduced to all the clinics in \revision{Pinggu county, a rural area in Beijing}, China~\cite{AIhealthcare}. \revision{Pinggu county has 18 villages, of which each has its own first-tier clinic. In total, there are 18 rural clinics in this county. In addition, this county also has 2 higher-tier care facilities (i.e., general hospitals). By the time this study took place, these clinics have deployed the ``Brilliant Doctor '' AI-CDSS system for six months. During the summer of 2019, we contacted all 18 first-tier clinics for this research, and 6 of them accepted our request to conduct field work on premise. The presidents of these six clinics coordinated the study by helping us recruit clinicians as study participants and secure IRB approvals.}
We collected data using ethnographic methods such as in-situ observations, semi-structured interviews, and contextual inquiries. In this section, we first present details related to data collection and analysis. Then, we briefly describe the user interface of the Brilliant Doctor system, followed by an overview of the research sites. 

\subsection{Research Participants}
Our participants included 22 clinicians (9 females, 13 males) from six different rural clinics (Table~\ref{tab:interviewee-table}). The types of training as well as years of experience varied across the participants. For example, their years of experience range from 8 to 32 years. The majority participants are physicians and surgeons with expertise in western medicine, whereas a few of them (n = 4) are specialized in Traditional Chinese medicine (TCM). \revision{Note that we did not interview patients in this study, as the directly-targeted users of AI-CDSS system are clinicians. Therefore, we only focused on clinicians in this study.}

\subsection{Data Collection}
\subsubsection{In-Situ Observations}
We spent a total of 10 days, each day with 4 to 6 hours in the clinics to study how AI-CDSS is used by clinicians in real world settings. Observations focused on different aspects of clinicians' work, including the artifacts, types of activities, and the use of systems. We collected a variety of materials, including observational notes, photos of the artifacts and systems. After each observation session, field notes were transcribed into an electronic observation log that included detailed descriptions and reflections of what was observed. All photos were taken with consent from clinicians and patients.

\subsubsection{Semi-structured and Contextual Interviews}
We conducted formal, 40 minutes to 1 hour long semi-structured interviews with 22 participants. All of our interviews were conducted in Chinese by native-language speakers in the research team. All the interviews were audio recorded with the permission from the participants. The interviews focused on work responsibilities, educational background and experience, user experience and perceptions of AI-CDSS, and concerns or barriers regarding the use of AI-CDSS in practice. In addition to the semi-structured interviews, we also conducted contextual inquiries during the observation. We \revision{asked questions about certain work practices or artifacts} that we found interesting or did not understand, but only after the clinicians completed their task. 

\begin{table}[]
\caption{Interviewee Information Table (Female = F, Male = M, Years of Practice = YOP )}
\begin{tabular}{cccc}
\hline
\textbf{Participants} & \textbf{Gender} & \textbf{YOP} & \textbf{Specialty} \\ \hline
I1 & F & 20 & Physician \\
I2 & M & 20 & Surgeon \\
I3 & M & 6 & Surgeon \\
I4 & F & 8 & Physician \\
I5 & M & 20 & Physician \\
I6 & M & 8 & Physician \\
I7 & M & 8 & Surgeon \\
I8 & F & 20 & Physician \\
I9 & F & 20 & Physician \\
I10 & F & 22 & Physician \\
I11 & M & 23 & Surgeon \\
I12 & F & 32 & Traditional Chinese Medicine \\
I13 & F & 24 & Physician \\
I14 & M & 28 & Surgeon \\
I15 & M & 20 & Traditional Chinese Medicine \\
I16 & M & 10 & Surgeon \\
I17 & F & 30 & Traditional Chinese Medicine \\
I18 & M & 8 & Physician \\
I19 & M & 26 & Physician \\
I20 & M & 10 & Traditional Chinese Medicine \\
I21 & M & 10 & Surgeon \\
I22 & F & 30 & Physician
\end{tabular}
\label{tab:interviewee-table}
\end{table}

\subsection{Data Analysis}
We used an open coding technique to analyze the observation and interview data. We first reviewed the electronic observation log and interview transcripts to get an overview of the context. All the transcripts are in Chinese, and the coding analysis is in Chinese too. We only translate the \revision{codes and quotes being} used in this paper. \revision{In the subsequent stage, we transferred data into Nvivo for further analysis. Two authors of the paper, who are fluent in both Chinese and English, conducted an iterative open coding process to identify themes, sub-themes, and representative quotes. Each coder independently coded a few transcripts in the first round, and then they discussed the codes to develop a codebook. In the second round of analysis, they used the codebook to analyze the rest of transcripts.} Our coding focused on work practices, patient-clinician interactions, types of tools used, and how and when AI-CDSS was used. After the second round of coding, we identified major themes describing the \revision{facilitators and barriers} of using AI-CDSS in practice, as well as suggestions to improve the design of AI-CDSS. This step was followed by identifying representative quotes to support the claims. We also performed content analysis of the photographs taken during fieldwork. This analysis provided additional contextual information, complementing the analysis of interviews and observation notes.

\subsection{The AI-CDSS System: ``Brilliant Doctor''}

\begin{figure*}[hptb]
  \includegraphics[width=\textwidth]{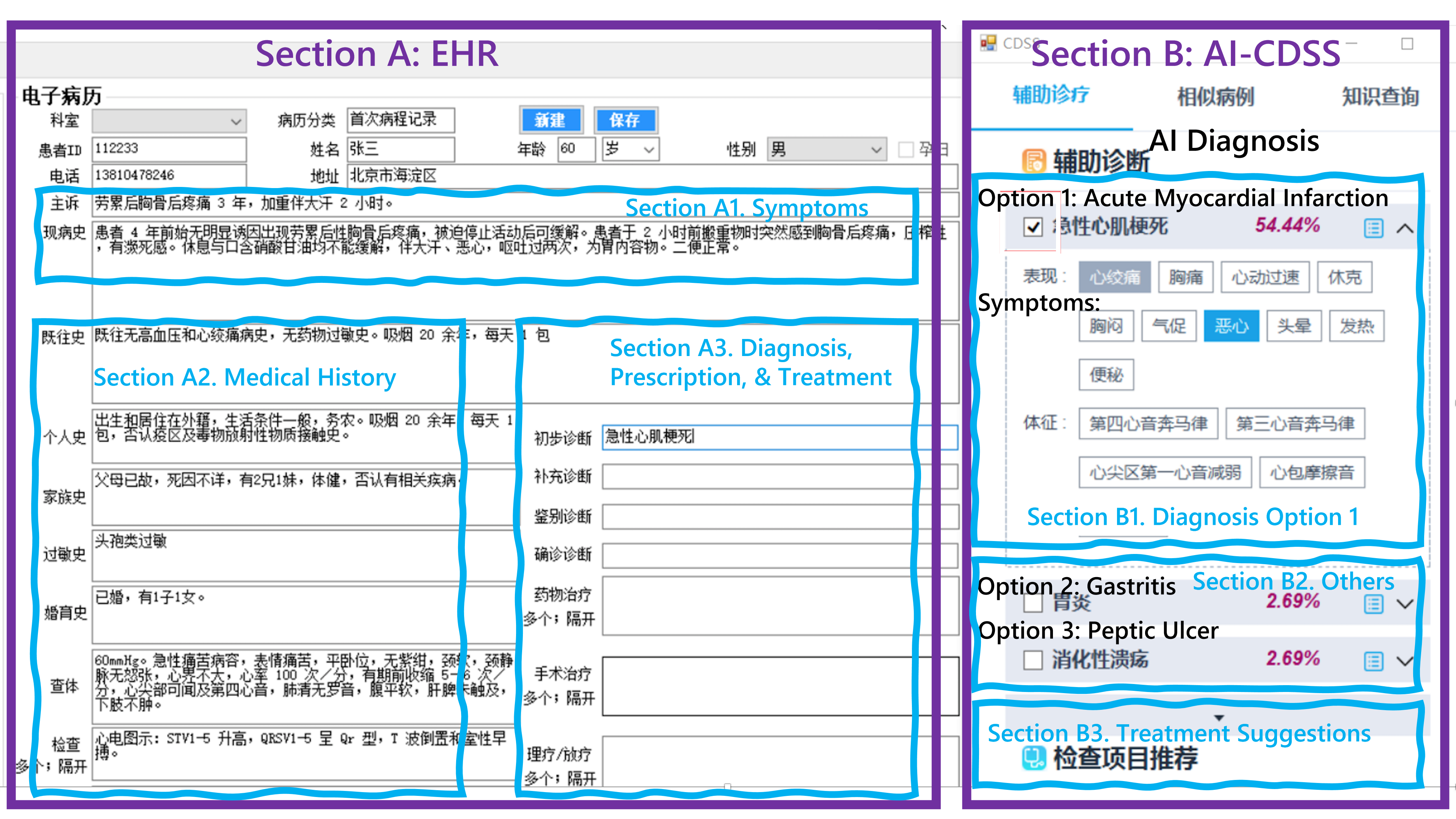}
  \caption{An illustration of the Brilliant Doctor system (to the right as a pop-up window) and the EHR system (to the left with gray backgrounds) we studied in this paper. The patient identity is anonymized with fake name and fake ID numbers. The EHR system has a patient's background, current symptoms (with patient's description and doctor's diagnosis), medical history, and other relevant information (e.g., allergy). The AI-CDSS system update its recommendation in realtime and suggest 3 options in the chart: Acute myocardial infarction with 54.44\% possibility, gastritis with 2.69\% possibility and peptic ulcer with 2.69\% possibility. In the pop-up window, many of the boxs are gray with one blue box, those indicates the particular symptom feature why AI made such suggestion. A full translated version of the screenshot is in Appendix.}
  \label{fig:cdss}
  \Description{An illustration of the Brilliant Doctor system (to the right as a pop-up window) and the EHR system (to the left with gray backgrounds) we studied in this paper. The patient identity is anonymized with fake name and fake ID numbers. The EHR system has a patient's background, current symptoms (with patient's description and doctor's diagnosis), medical history, and other relevant information (e.g., allergy). The AI-CDSS system update its recommendation in realtime and suggest 3 options in the chart: Acute myocardial infarction with 54.44\% possibility, gastritis with 2.69\% possibility and peptic ulcer with 2.69\% possibility. In the pop-up window, many of the boxs are gray with one blue box, those indicates the particular symptom feature why AI made such suggestion. A full translated version of the screenshot is in Appendix.}
\end{figure*}

\revision{The AI-CDSS system, namely ``Brilliant Doctor''\footnote{ https://01.baidu.com/cdss-basic.html}, is integrated with the existing EHR system in the clinics. Its user interface (UI) is merged with and placed next to the EHR window.} As illustrated in Figure~\ref{fig:cdss}, the integrated system has two main sections: the EHR window (Fig.~\ref{fig:cdss}, Section A) and the AI-CDSS window (Fig.~\ref{fig:cdss}, Section B).

The EHR interface has a structured text input design. The patient's demographic information can be documented at the top. Other three major sections of the EHR system include fields for doctors to enter the patient's symptoms and chief of complain\revision{t} (Fig.~\ref{fig:cdss}, Section A1), medical history and allergies  (Fig.~\ref{fig:cdss}, Section A2), as well as diagnosis, prescription, and treatment plan (if applicable) (Fig.~\ref{fig:cdss}, Section A3).

The interface of ``Brilliant Doctor'' is located to the right of the EHR system (Fig.~\ref{fig:cdss}, Section B). Its primary functionality is recommending a list of diagnostic options based on the patient medical context. Each diagnostic suggestion comes with a confidence score, and the prediction is produced and updated in real-time. For example, in Fig.~\ref{fig:cdss} Section B1, ``Brilliant Doctor '' suggests that this patient may have Acute Myocardial Infarction with a confidence score of 54.44\%. Its second and third suggestions are Gastritis, and Coronary Heart Disease, respectively  (Fig.~\ref{fig:cdss}, Section B2). \revision{Besides recommended diagnoses, AI-CDSS also suggests treatment plans, appropriate laboratory tests, and medications. These suggestions are organized into different sections which can be found by scrolling down}  ( Fig.~\ref{fig:cdss}, Section B3).


By design, this integrated system provides \textbf{two interaction mechanisms} for clinicians to use:
\begin{itemize}
    \item The system preserves clinicians' existing way of using EHR, which allows users to type in all the information (e.g., experienced symptoms, medical history) to the EHR UI in Fig.~\ref{fig:cdss} Section A. The AI-CDSS module actively parse the \revision{entered data and use that to highlight the symptoms in light blue color in Section B1. In the meantime, AI-CDSS is generating diagnostic suggestions which are constantly changing as new information is entered.} We call this approach as ``\textbf{Type-In}''.
    \item The second way of interaction requires clinicians to manually click through a list of pages to provide input to symptom-related and medical history questions, as well as other screening questions (e.g., diet and emotional state). This approach is designed to replicate the textbook workflow of clinical reasoning. We term this approach as ``\textbf{Click-Through}''. Once the clinician finishes all the questions, \revision{AI-CDSS will make several predictions for the clinician to consider. When a suggested diagnosis is chosen, the AI-CDSS system will} automatically generate text content to populate the EHR form in Fig.~\ref{fig:cdss} Section A.
\end{itemize}


In addition to its primary diagnosis recommendation function, ``Brilliant Doctor '' has two other primary features. One is retrieving and showing cases that are similar to the current patient, which helped clinicians learn more about possible symptoms and effective treatments. Another is a medical information search engine, which allows clinicians to look up up-to-date information related to a specific disease, a medicine, or a treatment option. This feature was designed as a on-premise ``Medical Wikipedia'' for clinicians to use, because all the EHR and AI-CDSS systems in these clinics are disconnected from the Internet for security purpose.





\subsection{Research Site}

\begin{figure*}[htpb]
  \includegraphics[width=\textwidth]{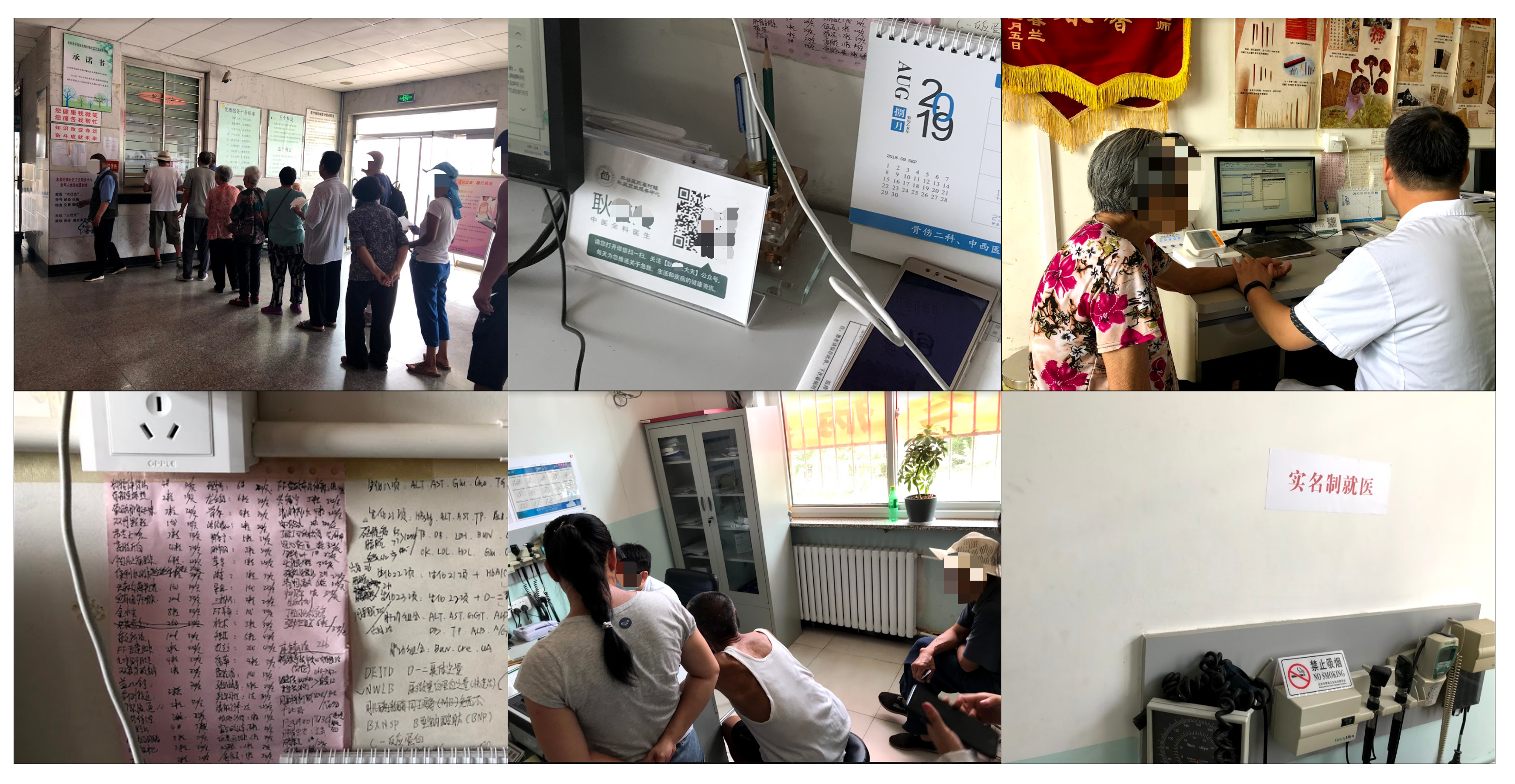}
  \caption{Photos from the Research Sites. From top left to bottom right, the first row shows 1) patients are waiting in line to pay for copay before seeing a doctor; 2) \revision{a doctor prints his name and Wechat QR code on a name tag so that the patients can ask him questions outside working hours;} 3) \revision{a TCM} doctor takes the pulse while typing into and interacting with the EHR system. The second row shows 4) \revision{personal notes about the medications that are allowed} to prescribe and in stock for that week in this clinic; 5) anxious patients are crowded in a clinic room, and use the clinic room as a social place to chat with other patients; 6) \revision{two wall signs---``do not smoke'' and ``use your own insurance card''}.}
  \label{fig:pin.png}
   \Description{Photos from the Research Sites. From top left to bottom right, the first row shows 1) patients are waiting in line to pay for seeing a doctor; 2) \revision{a doctor prints his name and Wechat QR code on a name tag so that the patients can ask him questions outside working hours;} 3) \revision{a TCM} doctor takes the pulse while typing into and interacting with the EHR system. The second row shows 4) \revision{personal notes about the medications that are allowed} to prescribe and in stock for that week in this clinic; 5) anxious patients are crowded in a clinic room, and use the clinic room as a social place to chat with other patients; 6) \revision{two wall signs---``do not smoke'' and ``use your own insurance card''}.}
\end{figure*}

As mentioned previously, we conducted this field study at 6 clinics, which are part of the 18 clinics providing healthcare services in a rural region near Beijing, China (as shown in Fig.~\ref{fig:pin.png}). In this subsection, we introduce the study sites by situating in the Chinese national healthcare system.

China's healthcare service has a strict hierarchical infrastructure where hospitals and clinics are categorized into three tiers according to their size, location, specialties, and so on~\cite{xu2014aging,collective_responsibility_2018, ding2019reading}. Tier-1 or the first tier is the lowest level, and tier-3 or the third-tier is the highest level. 
First-tier health providers are small clinics that are often located in rural areas (e.g., villages) or inside resident apartment complexes. They are designated to provide basic services such as prevention care, first-aid medical treatment, prescription refill, and rehabilitation. A few of these clinics may have a small in-patient department, which usually has less than 100 beds. Second-tier and third-tier providers comprised of general hospitals that are often located in cities or large towns. 
Most of them also undertake education and research responsibilities. Such a hierarchical healthcare system leads to some unique characteristics of these rural clinics including our research sites: 

\begin{itemize}
    \item \textbf{Overflow of out-patient visits at rural clinics}. \revision{By the end of 2019, there are 888,000 first-tier clinics and only 34,000 higher-tier hospitals in China, suggesting that the rural clinics (like our research sites) are predominant in China~\cite{statisticalcommunique}}. Even so, the healthcare resources are still far from enough to serve the 1.4 billion people in China.  \revision{According to ~\cite{statisticalreport}, the doctor-to-patient ratio is 2.77 clinicians per 1,000 patients across China.} To prevent patients squeezing into the top-tier general hospitals for basic services, residents in rural areas are not allowed to seek medical services in a higher-tier facility without first visiting their nearby clinics. This healthcare policy results in an overwhelming amount of patient visits to rural clinics every day. For example, in the rural area we studied, there are only 18 first-tier clinics and 2 second-tier hospitals; \revision{but these providers need to serve} over 400,000 residents (best depicted in Fig.~\ref{fig:pin.png} (1)). \revision{The doctor-to-patient ratio in this region roughly aligns with the nationwide ratio in the report ~\cite{statisticalreport}.}
    \item \textbf{Specialized clinicians also need to take the responsibilities of general physicians}. Because of the imbalanced doctor-to-patient ratio in the first-tier clinics, \revision{all specialized clinicians are required to take certification exams for general physicians so that they can take care of out-patient visits to reduce general physicians' workload.} This is a role similar to primary-care physician in the U.S. system or the general practitioner in the U.K. system. This policy leads to some interesting \revision{work practices}. For example, all the rural clinics we visited have a Traditional Chinese Medicines (TCM) department. The TCM doctors also take patients from other departments, and prescribe regular medicines beyond TCM domain.
    \item \textbf{Healthcare policy imposes strict regulations on first-tier clinics.} These first-tier clinics often do not have a laboratory department so that they can not perform any laboratory tests or radiology imaging studies. Also, according to the Chinese healthcare policy, the in-clinic pharmacy is not allowed to stock \revision{certain types of medication, so it is often that patients are not able to buy prescribed medicine in the clinic. Given this limitation, clinicians create personal note to help them remember what medicine can (or can not) be prescribed}, as shown in Fig.~\ref{fig:pin.png} (4). Another more important restriction is that the clinicians in first-tier clinics are only allowed to diagnose a limited number of diseases and health conditions. For example, even if a clinician suspects a patient has hypertension, he/she is only authorized to document the symptom, such as ``high blood pressure and need further examination'', instead of putting down ``hypertension'' as the diagnosis into the EHR system. The patient has to be referred to a higher-tier facility for further examination.
    \item \textbf{The insurance policy in China also has its own unique characteristics}. For example, patients with insurance are still required to pay the co-pay fee before seeing a clinician (also shown in Fig.~\ref{fig:pin.png} (1)). \revision{The clinic can periodically} get reimbursement for the deductible part from the insurance company. But, if the diagnosis or the prescription triggers alert (e.g., over-use of antibiotics), the insurance company reserves the right to reject certain transactions, and consequently, the clinicians \revision{who are responsible for the incidents} need to pay out of their pockets. \revision{This medical insurance policy is consistent across China.} 
    \item \textbf{The patients in rural areas are often residents \revision{who are less literate and financially vulnerable}.} Thus, they do not understand highly technical medical terminologies, and many of them speak in dialects. Clinicians have to use simple terms to communicate with them. They may not follow rules and regulations so that more than one patients could crowd into the doctor office (Fig.~\ref{fig:pin.png} (5)), and they sometimes use others' ID card for a better insurance coverage (Fig.~\ref{fig:pin.png} (6)). Many of them are also very sensitive to the expenses of the treatment and medicine.
\end{itemize}

With these strict regulations, it is not surprising to see that clinicians in first-tier clinics face significant challenges serving such a large population with only limited resources. In recent years, several initiates have been implemented to improve the quality of healthcare service in China, such as heavily investing in HITs ~\cite{zhang2017constructing}. These initiatives promote the deployment of ``Brilliant Doctor'' system in first-tier clinics, including our research sites. It provides us with a great opportunity to understand the use of AI-driven CDSS in the context of rural clinics.

\section{Findings}
In this section, we first report how the rural clinical context imposes various challenges on the adoption of AI-CDSS. Then, we present the technical issues and usability barriers in the ``Brilliant Doctor'' case. Lastly, we describe our participants' perceived benefits of the AI-CDSS system, their opinions regarding the future of AI-enabled CDSS systesm, and a set of suggested design considerations for improving AI-CDSS.


\begin{figure*}
  \includegraphics[width=\textwidth]{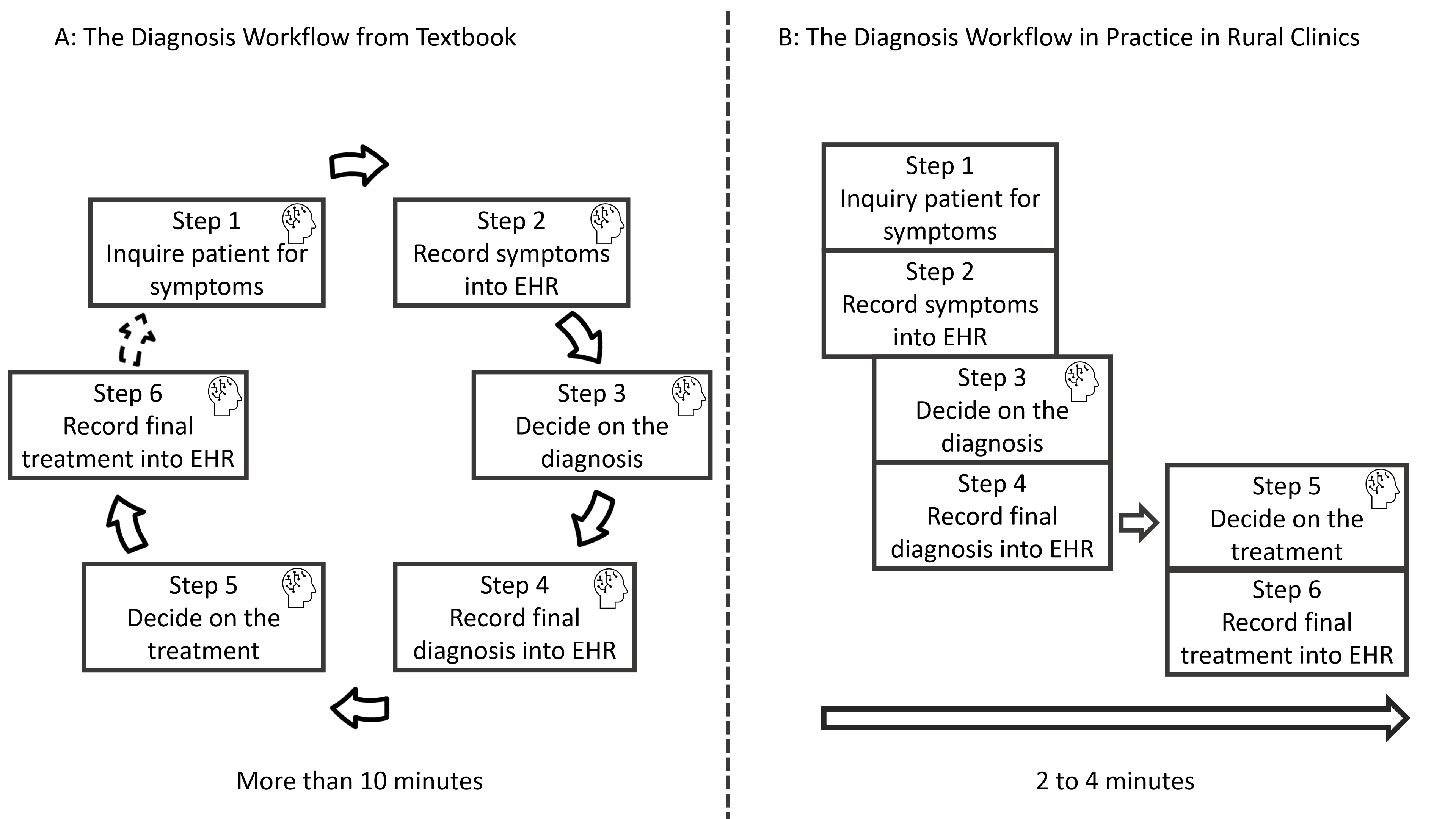}
  \caption{An abstracted view of the diagnosis workflow. A typical workflow consists of 6 steps: 1. inquiring patients to describe symptoms or chief of complaint, 2. documenting patient information into EHR, 3. making a diagnosis decision, 4. recording the diagnosis decision in the EHR system, 5. making a decision on laboratory test, treatment, or prescription, and 6. documenting the treatment plan in EHR. To the left, \revision{(A) depicts a ``textbook'' diagnosis workflow where the procedure of diagnosing a patient is clearly segmented. Each step is expected to take a few minutes so in theory a patient visit could take at least 10 minutes; To the right, (B) describes the actual workflow that the  clinicians follow to work around the high volume of patients in a rural clinic. There are lots of parralelization and the entire procedure takes ``about 2 to 4 minutes''. AI-CDSS was designed to support the ``textbook'' diagnosis workflow (A).} }
  \label{fig:process}
  \Description{The two workflows are placed on the left and right sides respectively. ``Textbook'' diagnosis workflow is on the left. It is presented in the form of a circle. It starts from step1 and points to the next step in turn. Step6 points to step1 again. The arrow that step6 points to step1 is a dotted line. The actual workflow is on the right. Step 1 and step 2 are in the same stage. They point to the second stage, in which step 3 and step 4 are carried out simultaneously. Then they point to the third stage, in which step 5 and step 6 are carried out simultaneously.}
\end{figure*}

\subsection{The Contextual Challenges to Adopting AI-CDSS in Rural Clinics}

\subsubsection{An “impossible” volume game}
As described in the Research Site section, the first-tier clinics in China primarily deal with ‘walk-in’ patients and general healthcare needs. That is, patients can visit these clinics anytime without making an appointment. Therefore, a primary care physician in such settings has an extreme\revision{ly} high workload due to the high demand of healthcare services. For example, a participant reported that in these rural clinics usually a doctor need to take care of 75 patients in a typical day. 

Due to the high patient volume, it is very common to have multiple anxious patients waiting in the doctor office at the same time (Figure~\ref{fig:pin.png} (2) and (4)). As a consequence, often a doctor's work is subject to constant interruptions from various anxious patients in the office asking questions and looking for clarification. As such, physicians are left with little time and capacity to talk to patients and answer their questions, let alone use the computing tools which were also competing for their attention. As one participant explained:  

\begin{quote}
 ``We are getting the most visits near the end of the year, averaging 150 per day for two clinicians. If you use AI-CDSS for every patient, the line will pile up. So what we do is we use the template [of the EHR system] , it will take two minutes to diagnose a person [with chronic disease in most cases], including inquiry and giving prescription.'' (I4)
\end{quote}

The participants believe that ``Brilliant Doctor'' is designed to support the ``textbook'' diagnosis workflow, and they acknowledge such workflow is more rigorous and legitimate. As depicted in Figure~\ref{fig:process} A, for example, a typical workflow of patient consultation consists of soliciting patient' demographics or medical history information, then recording patient information in the EHR system, making diagnosis, recording diagnosis decision in EHR, determining appropriate treatment plans and medicine to use, and finally recording such treatment information in EHR. The AI-CDSS was designed to support each individual step from Step 2 to Step 6 in the workflow. For example, in Step 2, AI-CDSS is designed to allow clinicians to click through 50 or so screening questions for diagnosis support. Also, in Step 5 and 6, the user can simply click on one of the suggested treatment recommendation (Section B3 in Fig~\ref{fig:cdss}) and the content is automatically populated into the EHR system.

However, it is simply impossible for these clinicians to follow the textbook workflow. To workaround the high patient volume and the constant interruptions in these clinics, clinicians had to multitask. For example, they record symptoms and chief of complaint into the EHR system (Step 2) while talking with the patient to inquire their symptoms (Step 1) (Figure~\ref{fig:pin.png} (3)). When they finish typing the symptoms, they already make a decision for the diagnosis (step 3) so their hands do not stop but to record the diagnosis into EHR (step 4). The clinicians only stop when they discuss with the patients about which treatment plan or medicine would they prefer, since these patients are often financially vulnerable.

In order to generate suggestions, AI-CDSS needs detailed information from the EHR system. In theory, the clinicians are required to ask about patients medical history, check existing medicine usage and emotional status, and then document all such information in the EHR system. However, in practice, due to high patient volume and time limits, our participants disclosed that they often do not have time to ask these detailed questions and document them properly. They believe these questions are ``nice to have'' but it is ``impossible to ask them all’’, given the workload they already have in the clinic \footnote{Even if the clinicians did input all the information, there is a system interoperability problem which is explained in detail in the next section}. 

\revision{What exacerbates the problem is that} these rural clinics are also experiencing shortage of medical staff, especially nurses, who usually perform initial check-ups such as checking blood pressure, recording the medical history, and taking the patient’s temperature, etc. As such, it is impossible for the clinicians to \revision{capture sufficient information for AI-CDSS to} make accurate and comprehensive diagnosis and in turn, the clinicians just do not use the system often.

This multitasking and shortened workflow is illustrated in Fig~\ref{fig:process} B. The AI-CDSS system designers had followed the textbook workflow and envisioned that clinicians would take time to process each step. But the unique characteristics of rural clinics posed significant challenges in the uptake of AI-CDSS, leading clinicians to have rarely used the system in practice. Even when they were using the system, they used the ``Type-In'' interaction approach rather than the ``Click-Through'' approach, since the latter approach requires much longer time to complete a diagnose process, \revision{even though ironically the ``Click-Through'' approach was designed to make clinician's work easier and more efficient}. 



\subsubsection{80\% patients visit first-tier clinics for trivial cases.}

Participants report that there are three types of patients who usually visit these rural clinics. Firstly, there are patients with chronic diseases who regularly visit the clinics to refill their medications. The second type of patients have new symptoms and minor disease, and their disease can be treated in first-tier clinics. Finally, there are also patients with more severe and complex disease who need to be referred to higher-tier hospitals for further examinations; \revision{but the health regulations and insurance policies require such patients} to get a referral by visiting their local first-tier clinics first. Typically, \revision{ ``80\% of the patients''} who visit the clinics belong to the first or the third category that they either require a refill or a referral. These cases do not require the clinician to provide a detailed diagnosis every time, but the EHR records need to be completed. Therefore, to speed up patient consultation, the clinicians always customize and reuse EHR templates to fill patient information in a timely manner, but AI-CDSS is barely used for these cases:

\begin{quote}
 ``We have a lot of patients that simply need to refill the prescription for their hypertension. We don't use it [AI-CDSS] when patients just come to pick up refills because AI-CDSS asks us to enter the necessary information (e.g., medical history) for a medicine refill case but we don't have time to do so. I just directly open up the EHR template I created for medicine refill.'' (I19)
\end{quote}

In addition, since our research sites are first-tier community clinics, they are only capable of performing a limited number of laboratory examinations (e.g., none of the research sites have CT-scan equipment). They also have very limited medication resources in stock. However, AI-CDSS would suggest a variety of laboratory tests, and treatment and medicine options, which clinicians often can not prescribe. In this case, theses recommendations are often ignored by the clinician users. As one physician explained:
\begin{quote}
    ``Our clinics have very limited resources. We know what kinds of laboratory tests we can do here so we don't really check AI-CDSS recommendations. We can only perform some basic laboratory tests, such as blood test, urine test, or electrocardiogram (ECG) examination. That is it. For example, this senior patient [pointing to the patient] has a medical history of bronchitis. Recently, he had a flu. He is experiencing shortness of breath. It is possible that he has peneunomia or emphysema. To further investigate, he needs to do a CT scan or other advanced laboratory tests. AI-CDSS is capable of suggesting the tests this patient needs to take, but we can not do those here. So what I would do is referring him to a higher-tier hospital so he can take those tests.`` (I2)
\end{quote}

\subsubsection{The Interoperability Issue}
Another major issue was associated with \revision{the interoperability between AI-CDSS and} other healthcare information systems in the clinics. AI-CDSS needs different kinds of data input in order to generate an accurate and reliable diagnosis. For example, it needs patient information as well as laboratory test results for diagnostic suggestions. However, in its current form, AI-CDSS is not seamlessly integrated with the laboratory information systems. As such, the missing \revision{information about laboratory tests} could make the suggested diagnoses less accurate. In a similar vein, AI-CDSS is not synchronized with systems used by other departments in the clinic, such as the pharmaceutical system. Some prescriptions recommended by AI-CDSS are not available in their pharmacy, and what they have in the pharmacy sometimes are not updated in the AI-CDSS system. Another extreme case is that the AI-CDSS system seems to be designed for internal medicine department only. There is no medical code for classifying TCM diagnosis in the system; none of the TCM medicine is available in the recommendation either. Thus, TCM clinicians found the AI-CDSS system less useful. These limitations could create a lot of confusion when clinicians use AI-CDSS to prescribe medicine. 

\begin{quote}
 ``I would recommend integrating AI-CDSS with our systems, including our pharmaceutical system and laboratory notification system. For example, AI-CDSS suggests to do some laboratory tests. Ideally, I can just simply click on these suggested test options and the laboratory department should be able to receive my orders. Also, our pharmacy may not have some suggested medicine. If you can import our pharmacy's data into AI-CDSS, or just connect these two systems, then it will be extremely useful for us. And in turn, the usage rate of AI-CDSS will be significantly higher.'' (I8) 
\end{quote}

To deal with this challenge, many clinicians create hand-written personal notes of the medicine that are available in stock for the week, as shown in Fig~\ref{fig:pin.png} (4). And they manually update the list every week. Sometimes, if a clinician prescribes an out-of-stock medication, the in-house pharmacy will make a phone call to the \revision{clinician regarding this matter and ask the clinician to change the patient's prescription. The clinician will update his/her personal note afterward}.

\subsection{Challenges related to Usability, Technical Limitations, and Trustworthiness of AI }
\subsubsection{Usability Barriers of the AI-CDSS system}
A primary issue of AI-CDSS usability was that the system always ``pop up'' to occupy one third of the screen, whenever the clinician opened a patient's medical record in EHR. If the monitor's screensize is small, the floating window of AI-CDSS may block the access to some EHR features (e.g., data fields). This frustrated many participants. To workaround this issue, clinicians had to minimize it while it was not in use. 
\begin{quote}
 ``It will pop up when a new patient record is opened. But I don't need to use it every single time so you literally have to move it to the side. Even that, it will block something, such as the ``Save'' or ``Exit'' button. I don't know if it is caused by the screen resolution or something else.'' (I7) 
\end{quote}

Another related issue is the display of system features. That is, all the features are organized into one pop-up window (Figure~\ref{fig:cdss}, Section B). For example, a set of features such as ``Diagnostic Suggestion'', ``Physical Examination Recommendation'', ``Laboratory Test Recommendation'', ``Medicine Suggestion'', and ``Surgery Suggestion'', are grouped into one tab. However, this style of information architecture is problematic as a scroll box must be used to view and use multiple functionalities. During our observations, we were surprised to find out that a few participants did not know the existence of some features (e.g., Similar Cases), as those features were not easy to discover. This issue was also evident in the display of alerts. When there were too many alerts in the scroll down box, such as ``be caution to prescribing medicine A'', clinicians did not read through them at all.

\subsubsection{Technical Limitations of the AI-CDSS}
Technical limitations of ``Brilliant Doctor'' were frequently mentioned by our participants. For example, clinicians felt that the suggested three diagnoses were ``generally good'', but the top recommendation was not always correct. Therefore, these recommendations did not provide useful insights and usually took up too much time to be worthwhile. Also, the confidence score of each suggested option could be pretty low, such as 30\%, making clinicians not willing to trust the recommended diagnosis. The low accuracy may be caused by two reasons. First, due to the interoperability issues between AI-CDSS and other systems, AI-CDSS lacks effective way to gather sufficient information, and thus, its outputs may not be presented with complete and relevant patient information. Second, the current A\revision{I}-CDSS system is not capable of capturing subtle clues that can only be recognized by clinicians. These clues, however, may be vital in the diagnostic process: 

\begin{quote}
``The patient's facial expression or other hints might indicate the severity of illness, but AI-CDSS was not able to interpret that.'' (I3) . 
\end{quote}

\begin{quote}
 ``Can you make it more precise, more intelligent. [...] The same symptom can be associated with different diseases or illnesses, and different patients have different medical histories and unique characteristics. The system is protocolized and not able to provide personalized diagnosis. Recommendations triggered on symptom alone were often not quite accurate.'' (I5)
\end{quote}

Another issue is that some information, such as medicine information, presented by AI-CDSS is not accurate either. Very often, the information recorded in the system is not updated, as one physician explained: 
\begin{quote}
    ``The medicine description provided by AI-CDSS, such as how much to take, is not always accurate. Take 'tamsulosin' as an example, it is used to treat prostate issues. AI-CDSS says take one pill per day so I followed its guideline. But some older adults have very serious prostate problem, just taking one pill is not effective. Some of them decided to take two pills per day without consulting us, but they said it works well. Therefore, I went down to the pharmacy and checked the description of this medicine, and found that it says take one to two pills per day instead of strictly taking one.'' (I7)
\end{quote}

Similar to the TCM example in the previous sub-section, almost all of the surgeons complained that the system is not very useful. The root cause of this problem is that AI-CDSS can not fully accommodate surgical issues, such as wound or fracture. One surgeon voiced his frustration:
\begin{quote}
 ``For example, when a patient who has wound injury comes in, this system doesn't have a full coverage of those surgical issues. The symptoms it can suggest and recognize mainly related to internal medicine. The last time I found it useful was when a patient who had abdomen pain, but still it is not quite a surgical issue.'' (I3)
\end{quote}

There are some other technical limitations brought up by the participants. In particular, the AI-CDSS algorithm does not take into consideration the patient's individual information. For example, it is not aware of a patient's income or insurance information and thus, the algorithm recommended medicine options may not be affordable by the patient. Lastly, it is worth noting that the participants expressed concerns that the extensive use of AI-CDSS could lead to overly reliance on the system and may even de-skill clinician, which leads to our next point on trust in AI and professional autonomy. 

\subsubsection{Trust in AI-CDSS and Professional Autonomy}
\revision{All of our participants indicated} that clinical expertise of the healthcare providers can not be easily replaced by an AI system, as one participant stated---``diagnosis is not a formula''. The AI-CDSS system can make mistakes too, thus the medical advice given by AI-CDSS should always be verified by clinicians. After all, clinicians are accountable for the treatment of their patients: 
\begin{quote}
 ``Doctors and patients are friends, we usually have a good relationship. It is possible that the prescription we gave to the patient is not working. I'll just recommend them to go to higher-tier hospitals for further examination. They understand us too, it is not like I intentionally gave you a wrong medicine or made you to be uncomfortable. [...] But if the AI system [directly] gives him a prescription that is not working, or unfortunately it causes some adverse events. The patient must complain about it. And more importantly, there is an accountability issue in there. Who is responsible for that?''(I19) 
\end{quote}

Furthermore, the perceived trust of AI-CDSS was impacted by the \textbf{``Black Box''} nature of AI. That is, even the system gives a correct recommendation, how it works inside the system, such as the rationale for how different features were considered by the system, remains opaque to clinicians. Therefore, clinicians had to spend time evaluating the recommendations given by the AI-CDSS system. One physician explained that the system often alerted the use of medicine but did not show all of the relevant information, such as why it gave such alerts. For example, some alerts are categorized as ``proceed with cautions'' and some others are ``critically forbidden'', but the system did not provide enough explanatory information as to why these alerts were generated and how severe they were. This finding is consistent with previous studies showing that AI systems need to provide explanations to increase human trust~\cite{yang2019unremarkable}.

Lack of training session for the system may be another amplifier to worsen the transparency and trust issue. Most of the participants reported receiving very limited training on how to use the system, thus they are not aware of many useful features. One clinician voiced his frustration:
\begin{quote}
 ``They didn't train us after the system was deployed. We had to figure out how to use the system on our own. No training sessions, no training materials, nothing. Nobody explained how to use the system, what features it has, where to find what, what feature is used for what task, etc.'' (I9) 
\end{quote}

In addition to trust issues, our findings also reveal that clinicians considered the AI-CDSS system could challenge their professional autonomy. For example, clinicians would like to rely on their knowledge and experience to diagnose a patient rather than being guided by the AI-CDSS system. That is another reason participants reported why the ``Click-Through'' approach was rarely used. Not only that clicking through different pages and answering all the questions is time-consuming, but also the system forces the clinician to follow its clinical reasoning process, eroding the natural flow of clinicians' work routines and diagnostic practice ~\cite{walter2008physician}. Our participants also expressed that they felt uncomfortable when they follow the detailed instructions and recommendations generated by AI-CDSS that advise them on what to do and not to do:
\begin{quote}
    ``Seems like the system was designed to replace doctors. It asks so many detailed questions one after another that could take forever to complete a diagnosis process. If you decide to use it, you have to provide answers to all of the questions. For example, you have to click through some questions that normally we don't ask, such as the patient's marriage status. And some other questions don't really make sense to us. Usually we chose to not ask the patient those suggested questions by AI-CDSS. But in order to get to the next page, we have to check off ``Not applicable''. Can you imagine how time-consuming and frustrating that is? So I just used this [click-through] feature a couple of times, and then never use it again.'' (I19)
\end{quote}

\subsubsection{Can AI Replace Human Clinicians?}
None of the participants believe AI can replace human clinicians. As aforementioned, there are accountability and legal issues with medical diagnosis and treatment. Because \revision{clinicians are responsible for the final decision and patient outcome}, and they do not fully trust AI-CDSS. Also, all participants argued that many patients from these rural areas do not know how to use computers. In addition, the AI-CDSS system user interface is full of medical terms and technical jargon. They see no way that their patients can directly interact with such an AI-CDSS system if it were an ``AI doctor'', even in the future.
\begin{quote}
    ``You will have to stay in medical school for 3 years in order to understand this [AI-CDSS UI].'' (I10)
\end{quote}

Some participants phrased that the diagnosis process is a highly interactive, communicative, and social event. They do not believe AI can ever replace human clinicians because they think AI can never have such social bounding and interaction with the patient, or to be a ``friend to the patient'' (I19), as clinicians did in this rural area. The social interaction of the consultation process is a critical part along with the diagnosis and treatment. Being inside a clinical room and talking to a doctor, the patient feels the treatment is reliable and authentic.

\begin{quote}
   ``[Sometimes] taking the medicine is like a placebo. These patients just want to come to see a doctor and have a conversation. Your words might be more effective than the medicine.'' (I21)
\end{quote}

During the interview, many clinicians expressed their expectation of AI-CDSS to be redesigned as a \textbf{``Doctor's Assistant''} (I19), as opposed to trying to substitute doctors' work, or telling doctors what to do or what not to do. They were not against to have more AI supports in their workflow, but they suggested to prioritize the design of the AI system as an assistant to take care of the low level and non-critical tasks so that clinicians can spend more time on patient examination and diagnosis.

\begin{quote}
   ``If you design and put it in the lobby as a receptionist or information desk. It can take screening questions and suggest the patient to visit which department. I think it can work this way -- to do some preparation work for the diagnosis'' (I5)
\end{quote}

\subsection{Perceived Usefulness of AI-CDSS}
Despite all the challenges and barriers in the adoption of AI-CDSS, almost all participants expressed their positive perceptions of AI-CDSS and optimistic attitude towards its future. 
Most of the participants (18/20) believed that the AI-CDSS system is already supporting various tasks of their work. We identified four common scenarios for which the AI-CDSS system play a critical role: supporting diagnostic process, facilitating medical information search, providing an on-the-job training opportunity, and preventing adverse events.

\subsubsection{Supporting Diagnostic Process.}
Despite criticizing the AI-CDSS's functionality and user experience, participants actually considered AI-CDSS to be useful for some tasks in their diagnose and decision making workflow. For example, participants acknowledged that the system-generated diagnostic and treatment suggestions can be used as a reference to consider other possibilities and alternatives that otherwise they might have not thought about. This usage of AI-CDSS is highly appreciated by our participants, as clinicians in rural clinics do not often see complicated symptoms and they may have cognitive inertia and treat a new disease as something they are more familiar with. As our participants elaborated:

\begin{quote}
 ``It gives us a reminder. We don't see many types of diseases very often down here in our clinic, but if someone has some diseases that have complex causes or something we have little knowledge about, you can use the system to assist your diagnosis. In particular, it gives you several options that can help you think about alternatives and take those into consideration. If I diagnose a patient by myself, I might misdiagnose him/her because I may overlook some aspects. Now we have this [AI-CDSS], so I can reference to the suggested diagnoses.'' (I6)
\end{quote}

\begin{quote}
``Because the system provides several diagnosis suggestions, it is a huge help for expanding our thoughts. In particular, the AI-CDSS system could facilitate the diagnosis of uncommon disease for new patients. I do feel it greatly improves our work.'' (I7)
\end{quote}

Many participants also explicitly pointed out that they frequently used the AI recommendations to verify their own diagnoses. In this way, AI-CDSS works as a reassurance mechanism in patient care. If they notice any discrepancy between their own diagnosis and the AI-CDSS's suggestions, they would further investigate the issue and check if something is wrong to prevent misdiagnosis. One participant explained this aspect:

\begin{quote}
 ``The diagnostic suggestions of the system allow you to verify your own diagnosis. More importantly, it makes you feel more confident if your diagnosis is consistent with the system suggestions.'' (I7)
 \end{quote}

\subsubsection{Facilitating Medical Information Search.}
The AI-CDSS system provides a search engine for clinicians to look up information about unfamiliar medicine or diseases. It works as an on-premise ``Medical Wikipedia'' to broaden clinicians' knowledge scope, and to save them a significant amount of time from searching and evaluating endless internet information. For example, before the implementation of AI-CDSS, clinicians used different mechanisms to find the necessary medicine information. Similar to Figure~\ref{fig:pin.png}(d), clinicians created paper notes with detailed medicine instruction (e.g., how many pills per day) and posted on the wall for quick reference. Or, they often had to call or pay a visit to the pharmacy department to check such instruction. Lastly, they also relied on their mobile phone to search information online. These alternates, however, are not as efficient as using the embedded search engine:
\begin{quote}
 ``We used the information search engine very often to look up medicine information. The reason is that there are a variety of medicine and the information about each medicine (e.g., side effects, use instructions) keeps updating, which makes it nearly impossible to remember all of that. Before the implementation of AI-CDSS, we had to create an excel sheet to document all the information and used that as a ``cheat sheet'' when seeing a patient. But now we just use the search engine embedded in AI-CDSS to look up medicine information in a timely manner. '' (I8)
 
\end{quote}

\subsubsection{On-the-Job Training Opportunity}
It is interesting to learn that some participants considered AI-CDSS as an opportunity for on-the-job training. For example, the ``Similar Case'' feature allows clinicians to look up similar patient cases to the case at hand, and these similar cases are all from top-tier research hospitals. This feature was perceived as not only useful for their diagnostic work, but also helpful for clinicians to self-educate and learn, as one participant noted: 
\begin{quote}
    ``I usually use the ``Similar Case'' feature when encountering a rarely seen case. It can not only augment your diagnosis, but also expand your knowledge pool. In particular, I like to open it up even when I am not seeing patients so that I can self-educate myself.'' (I19)
\end{quote}

Some of our participants echoed this statement and further noted that AI-CDSS may be particularly useful for less experienced junior clinicians. The diagnostic process is not only evidence-based, but also heavily relied on clinicians' years of experiences. When a junior clinician does not have much prior experience with some rarely seen symptoms, AI-CDSS could provide recommendations for him/her to follow, and in turn, the clinician can gain knowledge and experience:
\begin{quote}
    ``I think junior clinicians who just finished their residency can benefit a lot from using the AI-CDSS system. It is very possible that they don't have enough expertise to deal with some diseases or symptoms. They can follow the suggestions presented by AI-CDSS to come up with a diagnosis. During this process, they will also learn a lot and gain experience.'' (I22)
\end{quote}

\subsubsection{Preventing Adverse Events}
Last but not least, AI-CDSS can automatically detect and alert some common mistakes or risks in the treatment plan or medication prescription. For example, if a patient is pregnant but the clinician mistakenly ordered a computerized tomography (CT) scan in the EHR system, the system will automatically remind the clinician about the risk of performing CT scan on a pregnant patient. Another example is that the AI-CDSS can alert prescription errors or drug allergies. In this regard, AI-CDSS acts as a safe guard to alert clinicians about potential mis-treatments or mis-prescription of drugs, and to prevent adverse events and improve patient safety.

\begin{quote}
``Doctors are responsible for any misdiagnosis or misuse of medicine. But we may make mistakes and couldn't pinpoint all potential risks. If we make a mistake, it could adversely affect patient\revision{s}. The AI-CDSS system could help us avoid such serious issues. For example, if we prescribed both Dipyridamole and Aspirin Enteric-coated tablets to a patient, that would trigger an alert as these two medicine are not quite compatible and could lead to adverse drug reaction. Before AI-CDSS, we literally had to draw a table and join lines for side effects and incompatibility of medicine, and paste them on the wall for reference.'' (I6)
\end{quote}

\section{Discussion}
Participants believed that the AI-CDSS system has the potential to support many aspects of their work. However, many challenges inhibited effective and successful adoption and use of the system. Those barriers include difficulty in integrating with rural clinical context or with their diagnosis workflow in practice, lack of integration with other clinical systems, technical limitations, usability barriers, and issues related to the trustworthiness of AI-CDSS and clinicians' professional autonomy. Below we discuss how to address these issues.

\subsection{Design Implications for Adapting AI-CDSS to Local Context and Practice}
Our study is situated in rural areas of China where first-tier clinics are the primary healthcare resources for a significant \revision{number} of people. Echoing previous work~\cite{fraser2005implementing,shu2014ehr}, in our study, we found the ``Brilliant Doctor''\revision{ system was not used as extensive as predicted in large part due to the lack of consideration of the local context and work practices} that define Chinese rural clinics. For example, the low doctor-to-patient ratios \revision{in this context make it difficult for clinicians to sufficiently} address each patient's needs and questions. As described in our findings, these rural clinicians can only afford to spend a few minutes with each patient, otherwise the waiting patients who are also present in the doctor office may experience and even express their emotional distress and frustration. \revision{On the other hand, AI-CDSS takes more time to use. For example, the ``click-through'' interaction approach of the system usually requires at least 10 minutes for comprehensive data collection. This misalignment between AI-CDSS design and local context limits the clinicians' ability to use AI-CDSS in real-time. We propose that the AI-CDSS system can leverage the latest speech-recognition techniques to develop a voice-based user interface for hands-free data collection \cite{blackley2019speech}. This feature can automatically transcribe patient-provider conversations into text, parse the text, recognize and classify the contextual meaning of text utterances, and finally populate the corresponding EHR sections (e.g., mentioning of allergy in the dialog can be automatically added to the allergy history section in EHR). This voice-based interface has the potential to save clinicians from the time-consuming documentation during patient visits, allowing them to spend more time to interact with each patient. This design direction echoes the ``unremarkable AI'' design guideline to seamlessly and naturally integrate AI-based systems into the established workflow ~\cite{yang2019unremarkable}.}

Consistent with the findings reported by \cite{moxey2010computerized}, we noticed that recommendations generated by AI-CDSS are sometimes dismissed. One possible reason is that the system lacks personalized recommendations for individual patients. That is, the system failed to account for important contextual factors such as insurance policy. In this rural context, many patients may not have insurance (or their income is low) to cover certain types of medications. Thus, patients are very sensitive to the cost of the medicine or the treatment. It is a common practice in these rural clinics that clinicians need to discuss with the patient before deciding on the treatment or the prescription (as shown in Fig~\ref{fig:process}). Emerging work in the field of shared decision making highlights the importance of considering personal and contextual factors when determining care plans between healthcare providers and patients ~\cite{stiggelbout2012shared}. The AI-CDSS should therefore follow the guidelines of shared decision-making framework and take patients' social, cultural, and personal context into consideration in order to generate more personalized recommendations ~\cite{alden2014cultural}. \revision{To promote a self-adaptation capability for the system, system builders can leverage reinforcement learning algorithms, so that AI-CDSS's prediction errors or drifts can be detected and resolved even after the system is deployed with a pre-trained model in the real-world.}

Lastly, the first-tier clinics are different from higher-tier healthcare organizations (e.g., general hospitals). \revision{Unlike the few patients who need medical treatments or referrals, a majority of patients are existing patients who only need regular check-ups or medication refills for chronicle disease. But when they visit, they usually sign up (and pay) to see an expert or senior clinician---a practice that is completely unnecessary and consumes very valuable health resources in rural clinics. We believe that it will be helpful to deploy patient-facing, AI-powered self-serve triage systems, a technology that has been gaining momentum since the outbreak of COVID-19 pandemic~\cite{judson2020rapid}. For example, a self-serve triage system (e.g., in the form of kiosk or robot) could guide patients to enter their symptoms and reasons for the visit, and then direct them to different departments and clinicians based on their needs. For medication refills, patients only need to visit the pharmacy department. Similarly, regular check-ups can be assigned to the clinicians who are dedicated to these cases. By doing so, clinicians with specialists can be freed up to treat complex conditions that usually require longer time to allow for comprehensive examination. The self-serve triage system should be built and integrated with the existing AI-CDSS system so that the information collected by the triage system can be transmitted to the AI-CDSS system for further use. Clinicians therefore are able to instantly use AI-CDSS for decision support. When designing such self-triage systems, designers and developers should account for patients' needs and unique characteristics in the local context. For example, since our participants reported that a lot of their patients (e.g., senior citizens) have limited health literacy and technology proficiency, the system can use patient-friendly language or provide a voice-based interface to facilitate the use of the self-triage system.}

\subsection{Improving Technical Performance of AI-CDSS}

Our participants point out that the AI-CDSS system has many technical limitations which impede its ability to provide accurate diagnostic suggestions for effective interventions. For example, previous work found that unlike healthcare providers, AI-based CDSS is not capable of considering the ``whole patient''. A key example is that traditional Chinese medicine (TCM) clinicians usually rely on inspection, auscultation, question, and pulse-taking as the four major diagnostic methods to understand and diagnose the whole body's pathological conditions. \revision{These procedures, however, are not modeled in the AI-CDSS system, making the system less usable to TCM physicians.} Similarly, surgeons also questioned the usefulness of AI-CDSS in their domain given the system was primarily designed and developed for internal medicine use. 

It is well recognized that improving the algorithm and reducing the bias of training dataset (such as including more TCM and surgical clinical cases to train the models) are critical for increasing the usefulness and user-acceptance of AI-CDSS systems ~\cite{middleton2016clinical}. However, we also argue that as AI becomes more ingrained into day-to-day clinical practice yet its technical limitations can not be fully addressed in the near future, one approach that might be useful to increase the levels of user acceptance is informing front-line healthcare providers about the capabilities and limitations of AI-CDSS, such as which disease the system is better suited to address ~\cite{cai2019human}. 

All these considerations highlight the need of employing user-centered design approach to address the needs and concerns of clinicians as well as the barriers in integrating AI-CDSS to daily workflow. For example, \revision{the AI-CDSS system should be integrated with not only the EHR system but also other hospital systems (e.g., pharmaceutical system or laboratory test notification system), ensuring that the medications or laboratory tests recommended by AI-CDSS are consistent with their availability in these first-tier community clinics.}

\subsection{Designing Explainable, Trustworthy, and Cooperative AI-CDSS for Human-AI Collaboration in Clinical Scenarios}
\revision{Clinicians expressed that they did not understand how AI-CDSS generated recommendations for a given patient care. This issue may affect their trust toward the system. If such trust issues persist, only focusing on improving technical performance of the system is not meaningful since} AI recommendations that are delivered without appropriate context can be easily ignored by the intended users ~\cite{shortliffe2018clinical}. Our participants expressed the interest to know the scientific and clinical basis upon which the AI-CDSS system generates diagnostic predictions and medical advice, so that they are better positioned to evaluate the appropriateness of applying system recommendations to the patient. \revision{Yet existing AI-driven clinical systems still function as a ``black box'', making it difficult for clinicians to understand how and why a specific recommendation is made. Future studies should explore how to design explainable AI systems for healthcare providers.}

We also found that the perceived threat to professional autonomy could adversely affect clinicians' intention to use the AI-CDSS system. This suggests that there should be a clear boundary between what tasks AI can automate and \revision{what tasks clinician should lead. For example, our participants had little interest in letting} AI guide them through the clinical reasoning and diagnosis process. Instead of designing AI to replace or replicate a doctor's diagnosis work, AI may just act as a ``doctor assistant''. Some of the \revision{participants} have already started referring to the AI-CDSS system as an assistant, instead of a tool. 

\revision{Taken together, we argue that as more AI-based clinical systems are being developed and deployed in the near future, it will likely form a new paradigm of work in medical domain: clinicians and AI systems work together but with clear division of labor \cite{rehnmark2003effective}} to provide healthcare services to the patients. This new clinician-AI collaboration paradigm will extend the current human-in-the-loop AI design philosophy \cite{cranshaw2017calendar} and the mixed-initiative design approach~\cite{horvitz1999principles}. 
\revision{To support this new human-AI collaboration paradigm, the AI system should be designed to meet the explainable and trustworthy AI design guidelines; but more importantly,  the AI system should also be \textbf{``cooperative''} --- it can work together with human clinicians, fit into the local context, integrate with existing IT systems, and improve work productivity in the workflow. 
Last but not least, it is critical to define who is responsible for patient outcome in this collaborative framework. Health policies and initiatives should be enacted to address the accountability issue of AI-CDSS systems \cite{goodman2017european}.} 
Our future work will involve clinicians in the design process to explore what a cooperative ``doctor assistant'' AI-CDSS system looks like.

\subsection{Limitation}
This qualitative study only focused on a rural area in Beijing, China. Given the vast variations across regions in China, the result may not be generalizable to other rural areas in China or to other developing countries. However, our findings and design implications can provide valuable lessons and input for future studies looking into this research area. We also want to note that the algorithm of the studied system may not utilize the the-state-of-the-art AI techniques, which may affect the system technical reliability and accuracy. However, we do not believe an algorithm with higher accuracy can alternate the findings of our paper. Nevertheless, we urge machine learning and AI researchers to join the efforts and work together in AI-CDSS design and development from a socio-technical perspective.

\revision{We did not interview patients in this study, as the target users of AI-CDSS are clinicians. Therefore, we only focused on clinicians’ usage and opinions in this paper. This is a limitation as the patients are also potentially affected by the adoption of the system. Future research is needed to gather patient voices.}

\section{Conclusion}
In this work, we conducted an observational study and interviewed 22 clinicians at 6 clinics in a rural area in China to examine how clinicians use a recently deployed AI-CDSS system in practice. We found various challenges in users' adoption of the system, such as misalignment between the design of AI-CDSS and the local context and workflow, technical limitations and user experience issues, as well as trustworthiness of the system. We draw on these findings to discuss design implications for future AI-CDSS interventions to increase their uptake and adoption in rural clinics, and generalize the discussion for the broad Human-AI Collaboration paradigm. 

\section{Acknowledgments}
We thank all the interviewees for contributing their time and insights. We also thank the anonymous reviewers and editors for helping us to improve this work.


\bibliographystyle{ACM-Reference-Format}
\bibliography{main}

\clearpage 
\newpage
\section*{Appendix}
\begin{figure*}[hb]
  \includegraphics[width=\linewidth]{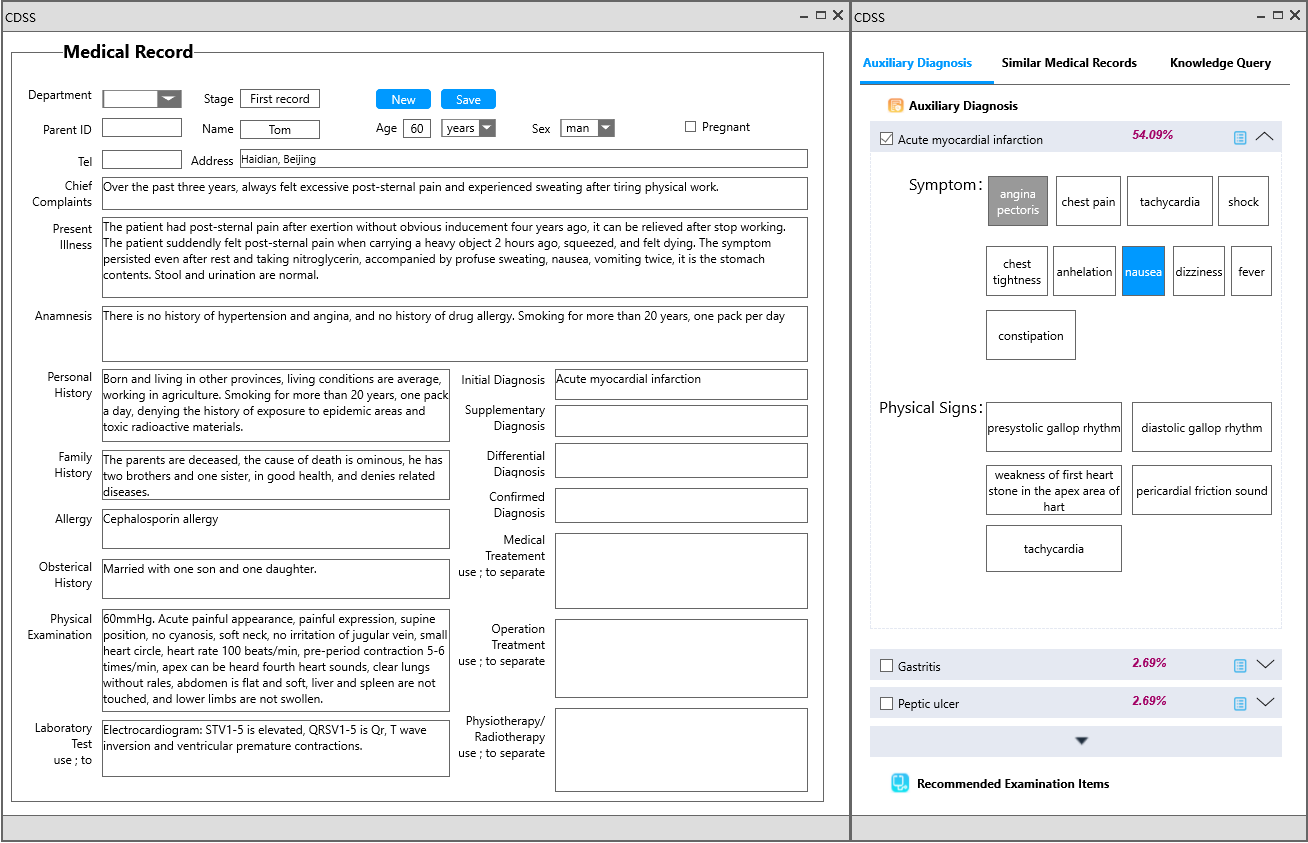}
  \caption{The English translation of the screenshot in Fig~\ref{fig:cdss} with the UI of EHR (left) and AI-CDSS system (right).}
  \label{fig:cdss_trans}
\end{figure*}

\end{document}